\documentclass[aps,pre,twocolumn,groupedaddress]{revtex4-1}
\usepackage{graphicx}
\usepackage{amssymb}
\usepackage{amsmath}
\usepackage[usenames,dvipsnames]{xcolor}

\begin{document}
\title{Steady Microfluidic Measurements of \\Mutual Diffusion Coefficients of Liquid Binary Mixtures}


\author{Anne Bouchaudy}
\author{Charles Loussert}
\author{Jean-Baptiste Salmon}
\email[]{jean-baptiste.salmon-exterieur@solvay.com}
\affiliation{CNRS, Solvay, LOF, UMR 5258, Univ. Bordeaux, F-33600 Pessac, France.}

\begin{abstract}
We present a microfluidic method leading to accurate measurements of the mutual diffusion coefficient
of a liquid binary mixture over the whole solute concentration range in a single experiment.
This method fully exploits solvent pervaporation through  a poly(dimethylsiloxane) (PDMS) membrane to obtain a steady concentration gradient within a microfluidic channel. Our method is applicable for solutes which cannot permeate through PDMS, and requires the activity and the density over the full concentration range as input parameters.
We demonstrate the accuracy of our methodology by measuring the mutual diffusion coefficient of the water (1) $+$ glycerol (2) mixture, from measurements of the concentration gradient using Raman confocal spectroscopy and the pervaporation-induced flow using particle tracking velocimetry.
\end{abstract}
\maketitle

\maketitle

\section*{Introduction}

Mass diffusivity in liquid mixtures is a key ingredient for designing
any  process involving mass transfer: mixing within chemical reactors, 
membrane-based separation processes~\cite{Strathmann:01}, 
drying of polymer solutions\dots~\cite{GU2005}\,
Current experimental techniques for measuring the mutual diffusion coefficient $D$ of a liquid binary system rely on the tracking of the relaxation of a concentration gradient within a cell, using for instance holographic interferometry~\cite{RuizBevia:85}
or spatially-resolved spectroscopy~\cite{Bardow:03}. In spite of their relevance, data sets  
reported in the literature still display significant discrepancies, mainly due to the difficulty of the corresponding experimental measurements. Indeed, molecular diffusion is a slow transport phenomenon which can be easily affected by any unwanted convective flux, possibly leading to the measurements of effective  diffusion coefficients~\cite{MACLEAN:01}. 
Moreover, current techniques only provide pointwise measurements thus requiring repetitive experiments when $D$ 
varies with concentration. To overcome this difficulty, several authors used model-based  diffusion experiments (with possible incremental model identification) to extract concentration-dependent diffusion coefficients in a single experiment, yet from time-resolved measurements of the relaxation of a concentration gradient~\cite{Bardow:03,Bardow:05}.

Microfluidics, as a toolbox for manipulating liquids at the nanolitre scale, provides outstanding opportunities for data acquisition in the field of chemical engineering, and particularly for diffusive transport~\cite{Chow:02,Jensen:99}.
Indeed, mass transport within liquids flowing in microchannels is perfectly described by mass balance equations  based on convection  and molecular diffusion only, because the microfluidic scale ($\leq 100~\mu$m) prevents from any unwanted buoyancy-driven convection and inertial effects~\cite{Squires:05,Stone:04}.
These unique features were successfully used by 
different groups to measure diffusion coefficients, using for instance 
co-flowing interdiffusing microfluidic streams~\cite{Dambrine:09,Hausler:12} or using time-resolved
measurements of the widening of a concentration gradient within a microfluidic chamber~\cite{Vogus:15}. However, such measurements cannot provide direct estimates of  mutual diffusivity over the 
whole solute concentration, without repeating tediously experiments at different concentrations.

\begin{figure}[htbp]
\begin{center}
\includegraphics{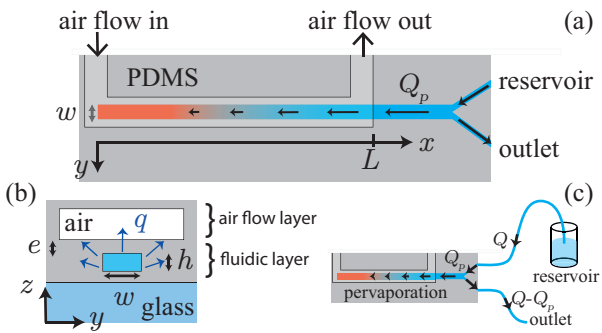}
\caption{{\small
(a) Schematic top view of the two-level chip. Water pervaporation from the fluidic channel drives a flow which concentrates 
solutes contained within the reservoir up to the tip of the channel, see arrows and Eq.~(\ref{eq:v}). The
colors show schematically the pervaporation-induced solute concentration gradient in the fluidic channel.
Typical dimensions are $h = 20~\mu$m, $w = 100~\mu$m, $L=10$~mm, and the pervaporation-induced flow rate is of the order of $Q_p = 1$--10~nL/min for pure water.
(b) Cross section of the device within the pervaporation
channel to evidence both the air flow layer and the fluidic layer. The blue arrows show water pervaporation across the PDMS membrane.  The typical thickness of the membrane is $e = 20~\mu$m.
(c) Schematic top view of the whole device. The microfluidic channel is connected to a feeding reservoir containing solutes.
A slight hydrostatic pressure drop imposes a flow from the reservoir up to the outlet at a rate $Q\gg Q_p$. This trick makes it possible to change rapidly the solutes which are concentrated at the inlet of the fluidic channel by plunging the tube into a different reservoir.}
\label{fig:Setup}}
\end{center}
\end{figure}

A few years ago, we developed original microfluidic tools for investigating waterborne complex fluids at the nanoliter scale. These tools harness water pervaporation through a poly(dimethylsiloxane) (PDMS) membrane, to concentrate in a controlled way, complex fluids confined within a microfluidic channel.
 The functioning of this technique is shown schematically in Fig.~\ref{fig:Setup}. 

%%%%%%%%%%%%%%%%%%%%%%%%% 	
%%%%%%%%%%%%%%%%%%%%%%%

Water pervaporation from a microfluidic channel 
(typical dimensions $h = 20~\mu$m, $w = 100~\mu$m, $L=10$~mm) through a thin PDMS membrane ($e \simeq 20~\mu$m),  induces a significant 
flow rate  within the channel of the order of $Q_p \simeq 1$-10~nL/min. This flow in turn convects the solutes contained in the reservoir towards the channel tip, where they accumulate continuously.
Over the past years, we used this microfluidic technique to screen 
phase diagrams of various systems ranging from  polymer and surfactant solutions to colloidal dispersions~\cite{Leng:07,Daubersies:13,Ziane:15}, but also to fabricate micro-materials with tailored architectures~\cite{Laval:16A,Angly:13,Laval:16}.

In the present work, we show that this microfluidic technique can also lead to accurate measurements of the mutual diffusion coefficient of a liquid binary mixture, and importantly to {\it continuous} measurements of this coefficient over the whole solute concentration using a {\it single} experiment.
 To illustrate our method, we focus on the well-known system water (1) $+$ glycerol (2), as different groups previously reported measurements of $D$, yet still with significant discrepancies~\cite{DERRICO2004,Ternstrom:96,Nishijima:60,RASHIDNIA2004}. 

The paper is organized as follows. We first explain in more details the mechanisms of microfluidic pervaporation in the case 
of an aqueous binary mixture solute (2) + water (1), and we show how such a technique can lead to estimates of its mutual diffusion coefficient $D$.
Then, we present the experiments performed along with concentration measurements using Raman confocal micro-spectroscopy, and velocity measurements using particle tracking velocimetry. 
We finally show that our technique leads to precise measurements of 
$D$ of the binary mixture water$+$glycerol over the whole range of solute concentration, and we compare our data to different measurements previously reported in the literature.

\section*{Experimental Technique}

Our microfluidic device is shown schematically in Fig.~\ref{fig:Setup}. It is a two-level
PDMS system sealed by a glass slide previously coated by a thin PDMS layer ($\simeq 15~\mu$m). The lower {\it fluidic} level
is composed of a microchannel with transverse dimensions $h = 35~\mu$m and
$w = 100~\mu$m, 
connected to a reservoir using a simple tube punched into the PDMS matrix and plunged into a vial, 
see Fig.~\ref{fig:Setup}. Microfabrication protocols of such chips can be found in Ref.~\cite{Laval:16A}.

An air flow of almost null humidity ($a_e \simeq 0$) is imposed within a large channel of the upper level of the PDMS chip, overlapping the fluidic channel over a length of
$L = 12~$mm. Pervaporation through the thin 
membrane ($e \simeq 15~\mu$m) separating the two channels  extracts water from the fluidic channel, thus inducing a flow $v(x)$ (m/s) within the lower channel. 
For pure water and for the geometrical features of our device, 
the pervaporation-induced flow rate $Q_p = (hw) v(L)$ is of the order of a 4~nL/min, leading to $v(L) \simeq 20~\mu$m/s
at the channel inlet, see later and Ref.~\cite{Laval:16}. 
When the pumped reservoir contains a binary solution   at a solute mass fraction $w_2^0$, 
the pervaporation-induced flow concentrates the solutes towards the tip of the channel, where they accumulate up to high concentrations, for solutes which cannot permeate through PDMS  such as glycerol~\cite{Lee:03A}.

A slight hydrostatic pressure from the feeding reservoir to the outlet imposes a flow $Q$ much larger than the pervaporation-induced flow rate $Q_p = 1$--10~nL/min.  This trick makes it possible to change rapidly the reservoir connected to the 
pervaporation channel simply by plunging the tube into a different reservoir (see later). 
Without the outlet and the flow at a rate $Q \gg Q_p$, i.e. with a single inlet connected to a tube plunged into a reservoir, the very low pervaporation-induced flow rate $Q_p = 1$--10~nL/min would  hinder the rapid draining of the feeding tube (typical volumes 1--10$~\mu$L) when plunged into another reservoir.

Schindler and Ajdari~\cite{Schindler:09} developed a theoretical model 
which describes this concentration process in the general case of binary liquid mixtures. 
This model yields the temporal evolutions of both the pervaporation-induced flow $v(x)$ and solute mass fraction
profile $w_2(x)$, using the following mass balance equations:
\begin{eqnarray}
&&(hw)[\partial_t \rho + \partial_x(\rho v)] = \rho^0_1 q_e (a(w_2)-a_e)\,, \label{eq:v} \\
&&\partial_t \rho_2 + \partial_x(\rho_2 v) = \partial_x (\rho D(w_2) \partial_x w_2)\,, \label{eq:convdiff}
\end{eqnarray}
where $\rho$ is the density of the mixture, $\rho^0_1$ the density of pure water, $\rho_i = \rho w_i$, $a(w_2)$ the water chemical activity 
at concentration $w_2$, and $D(w_2)$ the mutual diffusion coefficient of the mixture.
In Eq.~(\ref{eq:v}) which corresponds to the global mass conservation, the term $(a(w_2)-a_e)$ is  the local  driving force for pervaporation
across the membrane, and  $q_e$  is the pervaporation rate (per unit
length) in the case of pure water and a vanishing humidity $a_e = 0$. 
The neat control imparted by the microfabrication process 
ensures that $q_e$ is uniform over the channel length for a linear channel, see also experimental evidences of this feature in our 
earlier experimental works, in particular Ref.~\cite{Ziane:15}.
Note that $v(x)$ in the above equations corresponds to the mass-averaged velocity of the mixture (averaged over the transverse dimensions of the channel) defined as $\rho v = \rho_1 v_1 + \rho_2 v_2$,  where $\rho_i v_i$ is the mass flux of species $i$~\cite{Bird,Cussler}.

For most binary fluid mixtures, the volume of the fluids is unchanged by mixing. One can thus define 
unambiguously the volume fractions of species $i$ as $\varphi_i = \rho_i /\rho_i^0$, which further verify:
\begin{eqnarray}
\varphi_1 +\varphi_2 = 1\,. \label{eq:ideality}
\end{eqnarray}
In that case, it is more convenient to write the mass balance equations Eqs.~(\ref{eq:v}-\ref{eq:convdiff}) in the reference frame of the volume-averaged velocity $u=\varphi_1\,v_1 + \varphi_2\, v_2$ to remove explicitly the density $\rho$ from the model,
as done for instance in Ref.~\cite{Schindler:09}. Indeed, when Eq.~(\ref{eq:ideality}) applies, the volume-averaged velocity 
obeys $\nabla.u = 0$~\cite{Bird,Cussler}, and the above  equations take the simpler form:
\begin{eqnarray}
&&(hw)\partial_x u  = q_e (a(w_2)-a_e)\,, \label{eq:v_u} \\
&&\partial_t \varphi_2 + \partial_x(\varphi_2 u) = \partial_x (D(w_2) \partial_x \varphi_2)\,. \label{eq:convdiff_u}
\end{eqnarray}
In such a case, both velocities are related:
\begin{eqnarray}
v = u - (1/\rho_2^0 - 1/\rho_1^0)D \rho \nabla w_2\,, 
\end{eqnarray}
showing that a solute concentration gradient ($\nabla w_2 \neq 0$) induces mass convection ($v \neq 0$) even when $u=0$,
see e.g. Refs.~\cite{Bird,Cussler} for more insights. 
In the present work,  we prefer however to deal with Eqs.~(\ref{eq:v}-\ref{eq:convdiff}), in case our methodology would be applied to binary systems for which the volumes change significantly during mixing.    

Equations~(\ref{eq:v}-\ref{eq:convdiff})  as well as the implicit one dimensional approximation (or equivalently Eqs.~(\ref{eq:v_u}-\ref{eq:convdiff_u}) when Eq.~(\ref{eq:ideality}) is verified), have been discussed at length in Ref.~\cite{Schindler:09}
(see also Ref.~\cite{Salmon:10} for the dilute regime), and compared 
to experimental data obtained using various complex fluids~\cite{Leng:07,Daubersies:13,Ziane:15,Laval:16A,Angly:13,Laval:16}. 
The aim of the present work is not to discuss the pervaporation-induced concentration process in depth (see the above 
references for more details), but to show how to obtain ultimately a steady concentration gradient from which one can extract $D$ vs. $w_2$.

\begin{figure}[htbp]
\begin{center}
\includegraphics{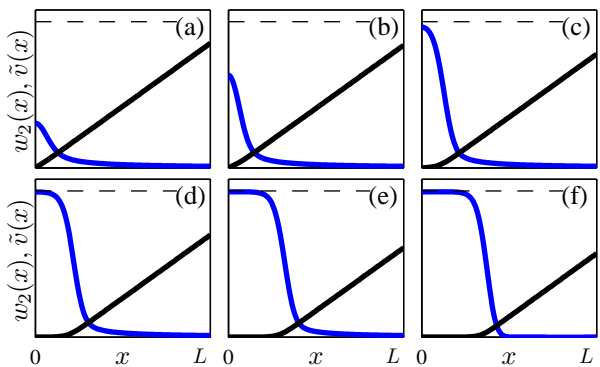}
\caption{{\small Schematic concentration process for a binary solution (blue: concentration profile $w_2(x)$, black: normalized pervaporation-induced flow  $\tilde{v}(x) = v(x) \tau_e /L$). $w_2(x)$ and $\tilde{v}(x)$ have been calculated using the numerical resolution of Eqs.~(\ref{eq:v}-\ref{eq:convdiff}), see text.  From (a) to (e): snapshots at  different increasing times. 
Between (e) and (f): the reservoir containing solutes is replaced by the reservoir containing pure water (i.e. $w_2(L) = 0$), and the concentration profile reaches a steady state in (f) (i.e. $\partial_t w_2 = 0$).
The dashed lines correspond to $w_2^\star \simeq 0.92$ given by $a(w_2^\star) = a_e$, with $a_e = 0.2$ in the case shown here (see text).}
\label{fig:Scenario}}
\end{center}
\end{figure}

To illustrate the expected concentration process, Figure~\ref{fig:Scenario}
shows the result of the numerical resolution of the above model, i.e. $v(x)$ and $w_2(x)$ calculated for different time scales $t$, in the case of the density $\rho(w_2)$ and activity $a(w_2)$ of the 
water (1) $+$ glycerol (2) mixture investigated in the present work, see Fig.~\ref{fig:Activity} later.
%%%%%%%%%%%%%%%%%%%%%%%%% 	
%%%%%%%%%%%%%%%%%%%%%%%
%For low concentration $w_2 \ll 1$, $a(w_2) \simeq 1$, and $a(w_2)$ drops to 0 for $w_2 \to 1$ as for most binary mixtures.
For the sake of simplicity, we solved the above equations with $D(w_2) = 4\times 10^{-10}~$m$^2$/s $= \text{cste}$, 
$w_2^0 = 0.01$, $a_e = 0.2$, and for the features of the microfluidic device investigated in the present work,
i.e. $\tau_e = (hw)/q_e = 600$~s, and $L=12$~mm (see later).
The reader is encouraged to refer to our earlier works~\cite{Salmon:10,Laval:16A}
and to Ref.~\cite{Schindler:09} for details about numerical resolutions with appropriate unitless variables, boundary conditions, and for a full discussion
of the convection-diffusion concentration process, including also analytical approximations. 
%With the assumption ,$\rho(w_2) = \text{cste}$ Eqs.~(\ref{eq:v}-\ref{eq:convdiff}) coincide with Eqs.~(\ref{eq:v_u}-\ref{eq:convdiff_u})
%with $u=v$.
%as the density of the mixture is constant. 
%For low concentration $w_2 \ll 1$, $a(w_2) \simeq 1$, and $a(w_2\to 1) \to 0$ as for most binary mixtures.
%With these assumptions, Eqs.~(\ref{eq:v}-\ref{eq:convdiff}) coincide with Eqs.~(\ref{eq:v_u}-\ref{eq:convdiff_u})
%as the density of the mixture is constant. 
%Eqs.~(\ref{eq:v}-\ref{eq:convdiff}) thus read:
%\begin{eqnarray}
%&&(hw)\partial_x v = q_e (a(w_2)-a_e)\,, \label{eq:v_s} \\
%&&\partial_t w_2 + \partial_x (w_2 v) = \partial_x (D(w_2) \partial_x w_2)\,. \label{eq:convdiff_s}
%\end{eqnarray}
%Initially, the reservoir contains a binary solution at a concentration $w_2^0 \ll 1$, and we thus impose the boundary condition 
%$w_2(L) = w_0$ at $x=L$. No-flux of solute at the channel tip leads to the second boundary condition $w_2 v(x) = D(w_2)\partial_x w_2$ at $x=0$. 
%(see the above cited references for more details).

At early time scales, the low 
concentration within the channel, $w_2(x) \ll 1$, hardly affects the chemical water activity and density, i.e. $a(w_2) \simeq 1$ and
$\rho(w_2) \simeq \rho_1^0$. Solutes accumulate owing to the pervaporation-induced flow at the channel tip, in 
a zone of size $p = \sqrt{D\tau_e}$, where the solute flux is dominated  by diffusion~\cite{Leng:07,Daubersies:13,Ziane:15,Laval:16A,Angly:13,Laval:16,Schindler:09}.
For the microfluidic device investigated in the present work $\tau_e \simeq 600$~s and typical molecular  diffusion coefficients, i.e. $D  = 4\times 10^{-10}$~m$^2$/s in the case shown in Fig.~2, yield $p \simeq 0.5$~mm.
For $x \gg p$, the concentration process is dominated by convection~\cite{Laval:16A,Angly:13,Schindler:09}, one has $w_2(x) \ll 1$, $\partial_t \rho \simeq 0$, and 
the velocity profile follows:
\begin{eqnarray}
v(x) \simeq \frac{1-a_e}{\tau_e} x\,, \label{eq:appV}
\end{eqnarray}
see Eq.~(\ref{eq:v}) and Fig.~\ref{fig:Scenario}(a).

Solute concentration increases at the tip of the channel towards $w_2^\star$ given by the local equilibrium $a(w_2^\star) = a_e$, because the decrease of the pervaporation driving force prevents  from further solute accumulation, see Eq.~(\ref{eq:v}).
In the specific case shown here, $a_e = 0.2$ leads to $w_2^\star \simeq 0.92$. 
As shown schematically in Fig.~\ref{fig:Scenario}(c--e), this plateau of $w_2 \simeq w_2^\star$  widens at longer time scales, and the velocity profile is shifted towards larger $x$ values within the channel. Far from the widening concentration gradient, concentrations indeed remain small $w_2 \ll 1$ and Eq.~(\ref{eq:v}) shows again that the 
slope of the pervaporation-induced velocity profile, $(1-a_e)/\tau_e$ see Eq.~(\ref{eq:appV}), remains constant.

A complete description of this scenario can be found in the above cited references.
In particular, we derived in Ref.~\cite{Salmon:10}, analytical relations which approximate the concentration field 
in the dilute regime, and in Ref.~\cite{Laval:16A} dedicated to the case of polymer solutions, analytical relations 
 to estimate the growth kinetics of the plateau $w_2 \simeq w_2^\star$ shown in Fig.~\ref{fig:Scenario}.
For the sake of brevity, we do not provide here these relations, but the reader is encouraged to refer to these earlier works to
estimate the different times shown in the panels of Fig.~\ref{fig:Scenario}, as 
a function of the operational ($w_2^0$, $a_e$), geometrical ($\tau_e$, $L$) and physical ($D$) parameters
of the experiments.

When the plateau $w_2 \simeq w_2^\star$ starts to grow within the channel (typically
for $t \simeq 20 \tau_e$ in the 
specific numerical simulation shown in Fig.~\ref{fig:Scenario}), one can replace the pumped reservoir containing solutes by a reservoir containing only pure water to obtain a steady concentration profile. Numerically, this steady state is obtained after imposing $w_2 = 0$ at the channel inlet at a given time
(precisely $t = 50 \tau_e$ in the 
simulation displayed in Fig.~\ref{fig:Scenario}).  
Solutes previously trapped within the channel reach, after a transient (of the order of a few $\tau_e$~\cite{Salmon:10}), a steady concentration profile 
($\partial_t w_2 = 0$)
\color{black}
given by the local equilibrium between convection and diffusion:
\begin{eqnarray}
&& w_2 v(x) = D(w_2) \partial_x w_2\,, \label{eq:equil} 
\end{eqnarray}
see Fig.~\ref{fig:Scenario}(f) and Eq.~(\ref{eq:convdiff}). 
This steady gradient can be used to estimate the mutual diffusion coefficient $D(w_2)$, because  
the shape of the profile $w_2(x)$
depends on $D(w_2)$ over the concentration range 0--$w_2^\star$.
More precisely, accurate measurements of the concentration profile $w_2(x)$ and $\tau_e$ 
can first lead to an estimate of the velocity profile $v(x)$ using the integration of Eq.~(\ref{eq:v}): 
\begin{eqnarray}
&&\rho v(x)  = \frac{ \rho^0_1}{\tau_e} \int_0^x \text{d}\tilde{x} (a(w_2(\tilde{x}))-a_e)\,. \label{eq:vint} 
\end{eqnarray}
Spatial derivative of the concentration profile $w_2(x)$ can then 
lead to values of  $D$ vs. $w_2$ using Eq.~(\ref{eq:equil}). 
Note that these experimental measurements require two thermodynamic inputs for estimating $D(w_2)$: 
the variations of $\rho(w_2)$  and  the water chemical activity $a(w_2)$.
Note also that the precise knowledge of the imposed humidity in the upper channel is not  required strictly as the concentration at the tip of the channel 
is expected to reach a plateau at $w_2^\star$ given by  $a(w_2^\star) = a_e$, see Fig.~\ref{fig:Setup}.
Note finally, that $\tau_e$ can be estimated using measurements of the velocity profile for positions $x$ beyond the steady gradient. In this region indeed, $w_2(x) = 0$ and Eq.~(\ref{eq:vint}) shows again that the velocity profile increases linearly with a slope 
$(1-a_e)/\tau_e$.

%In a previous work dedicated to a thorough investigation of the phase diagram of aqueous copolymer solutions using microfluidic pervaporation~\cite{Daubersies:13}, we used the methodology detailed above to estimate the mutual diffusion coefficient of this complex fluid. Unfortunately, these first measurements were only semi-quantitative as (i) $\tau_e$ and $w_2(x)$ were not measured with a very high accuracy, and more importantly (ii) $a(w_2)$ and $\rho(w_2)$ vs. $w_2$  were not known precisely for this specific mixture.
In the present work, we used the above methodology on a well-characterized system, water (1) $+$ glycerol (2), for which 
accurate data sets of both $a(w_2)$ and $\rho(w_2)$ exist in the literature. We first report accurate measurements of both the steady concentration profile $w_2(x)$ (precision $\pm 0.01$) and the evaporation time $\tau_e$ using Raman confocal
micro-spectroscopy and particle tracking velocimetry.
We finally show that these measurements lead to accurate values of the mutual diffusion coefficient $D(w_2)$  over the whole 
range of solute concentration.
 
\section*{Experimental procedure}

\subsection*{Materials and thermodynamic data}
Glycerol was purchased from Sigma Chemical Co.
(spectrophotometric grade, purity $> 99.5\%$) and it was used without further purification. 
For all our measurements, solutions were prepared by weighing using distilled water (milliQ water, 18.2~m$\Omega$ at 25$^\circ$C).

\begin{figure}[htbp]
\begin{center}
\includegraphics{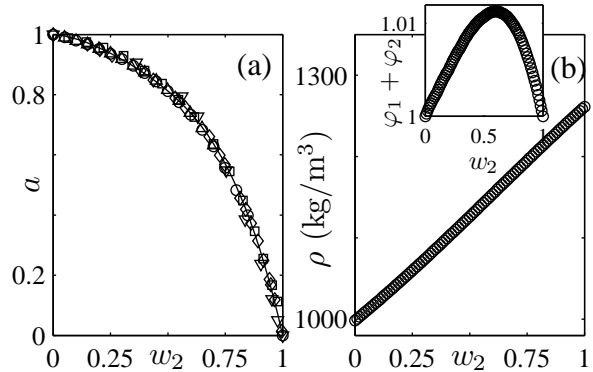}
\caption{{\small (a) Water chemical activity in water $+$ glycerol mixture vs. $w_2$. $\circ$ and $\triangle$  data from Ref.~\cite{NINNI2000} at 25 and $35^\circ$C resp. $\triangledown$  data  from Ref.~\cite{Zaoui2014} at $20^\circ$C.
$\square$ data from Ref.~\cite{MARCOLLI2005} at $25^\circ$C, $\Diamond$  data  from Ref.~\cite{Kirgintsev:62} at $25^\circ$C.
The continuous line corresponds to Eq.~(\ref{eq:act}).
(b) Density of the water $+$ glycerol mixture vs. $w_2$ at $20^\circ$C from Ref.~\cite{Glycerol}. The inset displays
$\varphi_1 + \varphi_2 = \rho_1/\rho^0_1 + \rho_2/\rho^0_2$ vs. $w_2$.} 
\label{fig:Activity}}
\end{center}
\end{figure}

Figure~\ref{fig:Activity}(a) displays several data sets of $a(w_2)$ measured by different groups
at several temperatures ranging from 20 to 35$^\circ$C~\cite{NINNI2000,MARCOLLI2005,Kirgintsev:62,Zaoui2014}. The variations of the chemical activity curves with the temperature are small, and all these data are well-fitted by the empirical relation:
\begin{eqnarray}
a(w_2) &=& (1-w_2)(1.6514 w_2^3 -0.2362 w_2^2 \notag \\ 
&+& 0.9542 w_2 +1 )\,, \label{eq:act}
\end{eqnarray}
with absolute deviations below $\pm 0.01$. We use the above equation in our method to compute the velocity profile $v(x)$ from 
the measurements of $w_2(x)$, see Eq.~(\ref{eq:vint}).
%%%%%%%%%%%%%%%%%%%%%%%%% 	
%%%%%%%%%%%%%%%%%%%%%%%

Figure~\ref{fig:Activity}(b) shows the density of water $+$ glycerol mixture $\rho$ vs.~$w_2$ at $20^\circ$C (precision $\pm 0.1$~kg/m$^3$) from Ref.~\cite{Glycerol}, and the inset displays $\varphi_1 + \varphi_2 = \rho_1/\rho^0_1 + \rho_2/\rho^0_2$ vs. $w_2$.
These data show that the  water $+$ glycerol mixture deviates from the ideal case described in Eq.~(\ref{eq:ideality}) by about only $\simeq 0.01$ at $w_2 \simeq 0.6$. This indicates that the volumes of this fluid  mixture do not change significantly during mixing, and we could 
have also safely used the reference frame of the volume-averaged velocity, see Eqs.~(\ref{eq:v_u}-\ref{eq:convdiff_u}), to extract $D$ vs. $w_2$. 
Nevertheless, we used in the following Eqs.~(\ref{eq:v}-\ref{eq:convdiff}), in case our methodology would be applied to binary systems for which the volumes change significantly during mixing.    

\subsection*{Concentration measurements}

We performed confocal micro-spectroscopy to get accurate measurements of the local concentration of 
glycerol in the microfluidic channel, using a Raman spectrometer coupled to an inverted microscope
(microscope Olympus IX71, Spectrometer Andor Shamrock 303i, laser   Coherent Sapphire SF with wavelength 532~nm). A confocal pinhole (100~$\mu$m) conjugated with the focal plane prevents from collecting excessive out-of-focus contributions.

\subsubsection*{Calibration}
To measure $w_2(x)$ within the chip, we first performed a careful calibration using vials containing water$+$glycerol mixtures at known concentrations. 
Raman spectra were acquired directly in the vials using a 20X~objective (Olympus, numerical aperture NA of 0.45), and typical experimental parameters are: acquisition time 30~s, slit 100~$\mu$m, and grating 600 lines/mm. The spectral range of these acquisitions 
is 1800-4500~cm$^{-1}$.

\begin{figure}[htbp]
\begin{center}
\includegraphics{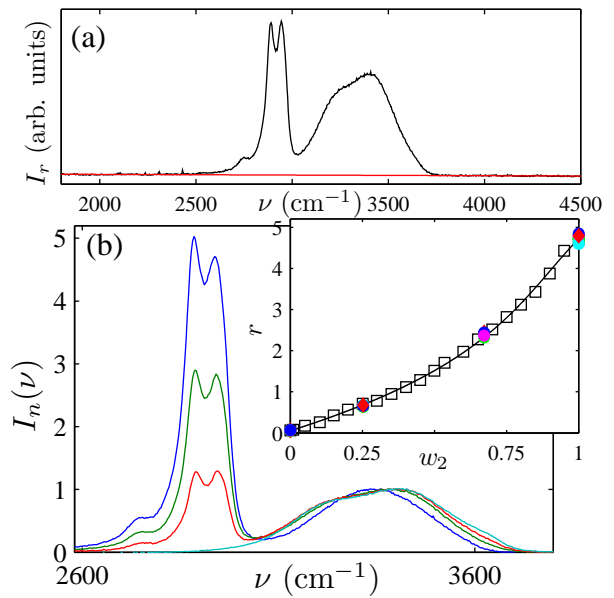} 
\caption{{\small (a) Raw spectrum $I_r(\nu)$ for a mass fraction $w_2 = 0.4993$ (black). The red line is an affine law which accounts for the baseline (see text).
(b) Normalized corrected Raman spectra $I_n(\nu)$ for mass fractions $w_2 = 0$; 0.4503; 0.7501; 1.0000.
The inset displays $r$ vs. $w_2$, $r$ corresponds
to the intensity of the second glycerol peak located at 2945~cm$^{-1}$ (black squares).
The continuous line is a fit by a 3$^\text{rd}$ order polynomial (see text). 
The (superimposed) colored symbols display similar measurements performed using different objectives to assess the precision of the measurements in vials (magnification 4X magenta, 10X green, 20X cyan, 60X blue). The red symbols correspond to on-chip measurements, see text.} 
\label{fig:Calib}}
\end{center}
\end{figure}

Figure~\ref{fig:Calib}(a) displays such a typical measured raw spectrum $I_r$ vs. $\nu$. Such spectra display a flat contribution in the spectral range 1800--2250 and 3850-4500~cm$^{-1}$ superimposed with the Raman contributions of water and glycerol. We first 
   corrected these raw spectra for their baseline estimated using a fit by an affine law in the spectral range $[$1800-2250~\&~ 3850-4500~cm$^{-1}$$]$. We believe that the baseline accounts for any wavelength-independent noise recorded by the spectrometer during the acquisition (e.g. electronic noise, contribution of the vial, etc.).  
Figure~\ref{fig:Calib}(b) now displays some corrected spectra zoomed in the range 2600-3700~cm$^{-1}$ for different glycerol concentrations. These data evidence 
a broad contribution in the region 3100--3600~cm$^{-1}$ due to the stretching and bending modes of the OH molecular bond. We used this contribution to normalize all the spectra by their maxima located in the range 3330--3400~cm$^{-1}$.
Note that the shape of this broad contribution changes with the glycerol content (e.g. their maxima shift from  3400 to 3340~cm$^{-1}$) thus probably pointing out the role of the local  composition on the OH vibrations.
 %%%%%%%%%%%%%%%%%%%%%%%%% 	
%%%%%%%%%%%%%%%%%%%%%%%

For $w_2 > 0$, spectra evidence two well-defined peaks at 2885 and 2945 cm$^{-1}$ corresponding to the glycerol contribution only. Note however that the measured intensity in this spectral range accounts for both water and glycerol contributions, as the two signals overlap. We defined $r$ as the maximum of intensity of the peak located at 2945 cm$^{-1}$. This peak, arbitrary chosen from the two, is  fitted by a local 2$^\text{nd}$ order polynomial fit (in the range 2930--2960~cm$^{-1}$) from which we extract its maximum. For $w_2 =0$,  we used the mean value estimated from the same polynomial fit of the remaining water contribution to estimate $r$. 
In the following, we refer $r$ to as a {\it ratio}, as it corresponds to the ratio that the Raman contribution of glycerol is to the broad contribution coming from the OH vibrations.

The inset of Figure~\ref{fig:Calib}(b) reports the ratio $r$ precisely estimated with this maximum of the second glycerol peak located at 2945 cm$^{-1}$.   
The curve $r$  vs. $w_2$ displays a well-defined shape which is nicely fitted by a 3$^\text{rd}$ order polynomial, with absolute variations below $\pm 0.01$. Lower order polynomials lead to poorly fitted data, whereas higher order polynomials
(4,5) were also tested without affecting the following estimates of mutual diffusion coefficients. 
Such a calibration allows us to estimate the glycerol mass fraction $w_2$ from 
the measurement of the Raman spectrum of an unknown mixture. 

To assess precisely the precision of 
such a calibration for estimating $w_2$, we performed, several months after the measurements shown with black squares in the inset of Fig.~\ref{fig:Calib}(b), similar measurements for four glycerol contents $w_2$. Each measurements were performed three times using different objectives (magnification 4X, 10X, 20X, and 60X, Olympus) and using the same 
optical configuration as above (note, however, that our custom-made optical setup has been re-aligned several times during this period). These measurements, see the colored points in the inset of
Fig.~\ref{fig:Calib}(b), fit perfectly with our previous calibration, and deviations from the 
3$^\text{rd}$ order polynomial lead to absolute variations below $\pm 0.01$ for the estimated $w_2$.

\subsubsection*{On-chip concentration measurements}

Concentration profile within the chip is obtained using a confocal configuration with the focal plane located at the channel center (along the directions $z$ and $y$, see Fig.~\ref{fig:Setup}(b)).
Typical experimental parameters are: objective 60X (Olympus, oil immersion, NA of 1.42), acquisition time 1~s, slit 200~$\mu$m, laser power at the focal plane $\simeq 25$~mW, and  grating
of 600 lines/mm.  The chip is displaced along the channel following the direction $x$, see Fig.~\ref{fig:Setup}(a), 
using a motorized stage synchronized with the Raman
acquisitions (M\"{a}rzh\"{a}user). 
The total duration of the Raman scan over an $x$-range of $\simeq 6$~mm is typically 20~min. 

\begin{figure}[htbp]
\begin{center}
\includegraphics{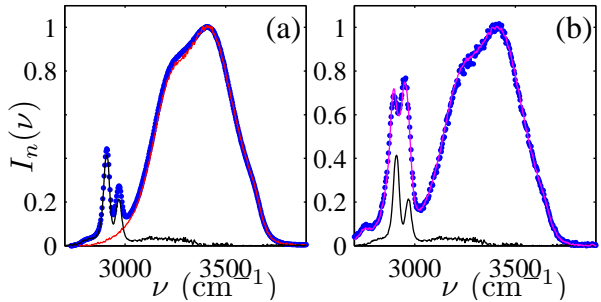}
\caption{{\small (a) Blue dots: Average (normalized) Raman spectra for $x>4.5~$mm. The red and black lines are respectively
the contribution of pure water and PDMS. (b) Blue dots: normalized 
Raman spectra measured at a specific position $x$, corrected  for the PDMS contribution (black line). The magenta curve is the closest Raman spectrum measured from a vial at a known concentration ($w_2 = 0.28$).}
\label{fig:SpectreMesure}}
\end{center}
\end{figure}

The baselines of the measured Raman spectra are subtracted as for the calibration. Far from the channel tip (i.e. for large $x$ values), one expects to measure pure water only, see Fig.~\ref{fig:Scenario}(f). However, our data evidence a significant contribution of the PDMS matrix despite the confocal pinhole (100~$\mu$m)~\cite{Everall:10}, see Fig.~\ref{fig:SpectreMesure}(a). 
We proceeded as follows to avoid this contribution and get accurate measurements of $w_2(x)$. We first averaged all the
spectra measured in the pure water region, i.e.  for positions $x > 4.5$~mm within the channel, see 
the spectrum with blue dots in Fig.~\ref{fig:SpectreMesure}(a). We then subtracted from this averaged spectrum, the contribution of pure water obtained at the calibration step (red line). The  remaining signal (black line) is consistent with the PDMS Raman spectrum  (two peaks located at 2910 and 2970~cm$^{-1}$) measured independently within the bulk of the PDMS chip (not shown). This PDMS signal (black line) is then subtracted identically from all the recorded spectra in the channel, i.e. for all $x$. Figure 5(b) displays
such a  PDMS-corrected spectrum at a given location $x$ superimposed with the PDMS contribution (black line). 
The position of the glycerol peak used to estimate $r$ ($\simeq 2945$~cm$^{-1}$) lies in between the two PDMS peaks, 
and the PDMS contribution is only of the order of $\simeq 0.1$ at $2945$~cm$^{-1}$. We finally extracted the ratio $r$ from these PDMS-corrected data, and from the measured $r(x)$ along the channel, we finally get $w_2(x)$ using the calibration curve displayed in the inset of Fig.~\ref{fig:Calib}(b).
%%%%%%%%%%%%%%%%%%%%%%%%% 	
%%%%%%%%%%%%%%%%%%%%%%%

To check the validity of the calibration despite the PDMS contribution, and to assess the precision of the on-chip measurements,  we performed the following experiments. We first made a straight microfluidic channel (with transverse dimensions $h = 33~\mu$m and $w = 100~\mu$m) within a thick PDMS matrix to avoid pervaporation ($\simeq 1$~cm). We then flowed water and acquired the corresponding Raman spectrum at a focal plane centered within the channel. Again, these data display a PDMS contribution despite the confocal pinhole, and its contribution is estimated as above using the subtraction of the water Raman spectrum acquired at the calibration step. We then flowed different water/glycerol solutions at known $w_2$,
and measured the corresponding Raman signals (without changing the optical configuration). We finally performed 
the same signal processing as above to estimate $r$ (baseline correction, identical PDMS subtraction). The corresponding 
data, reported in Fig.~4(c) with red symbols, show that the deviations from the estimated $w_2$ using the calibration curve are below $\pm 0.01$. These additional measurements help us to claim that our calibration lead to estimates of $w_2$ with a precision of $\pm 0.01$, even on-chip and despite the out-of-focus contribution of PDMS.

\subsection*{Microfluidic experiments and measurements of the evaporation time}
Experiments were performed using the microfluidic chip displayed in Fig.~\ref{fig:Setup}.
Since the transverse dimensions of the microfluidic channel are small, buoyancy-driven convection induced by small temperature gradients is negligible, and a strict temperature control is unnecessary. All the experiments were performed at room temperature $T = 20 \pm 0.5^\circ$C. 
  
	The microchannel is filled initially from a reservoir of a water$+$glycerol solution at a mass fraction $w_2^0 = 0.05$.
After a delay time of 15~min, we simply plunged the 
feeding tube into a reservoir containing only pure water. The small hydrostatic pressure difference between
the reservoir and the outlet imposes a flow at a rate $Q \gg Q_p$, see Fig.~\ref{fig:Setup}, which rapidly sets the 
solute concentration at the inlet of the fluidic channel at $w_2 = 0$. The pervaporation-induced solute flux is therefore 
zero, and one expects the build-up of a steady concentration gradient for the solutes previously trapped within the channel.

Concentration measurements are performed at a time long enough to get a steady concentration profile ($t > 3$~h, Raman measurements last about 20~min).
After complete Raman measurements, the reservoir of pure water is exchanged with a reservoir containing a dilute aqueous dispersion of fluorescent tracers 
(Fluorospheres Invitrogen, diameter 1~$\mu$m, concentration 0.002\% solids). The inlet of the pervaporation channel is thus fed again by fluorescent tracers which are then convected within the main channel, and we use particle tracking velocimetry to measure the velocity $v(x)$ at several $x$ positions.
More precisely, we measure series of images (typical duration 30--40~s) using an inverted fluorescent microscope (Olympus IX71) and a high numerical aperture objective (Olympus 60X, oil immersion, NA of 1.42) at a focal plane located at $z=h/2$ within the microchannel, using a s-CMOS camera (Hamamatsu, Orca Flash 4.0LT) . Significant threshold on the measured fluorescent intensities allows us to select in-plane tracers (focal depth $<1~\mu$m, maximal measured intensities $\simeq 2500$, 
dark intensity $\simeq 50$, threshold  500).
Particle identification and tracking is performed using the Particle Tracking Code developed by Blair and Dufresne~\cite{PTV}.

\begin{figure}[htbp]
\begin{center}
\includegraphics{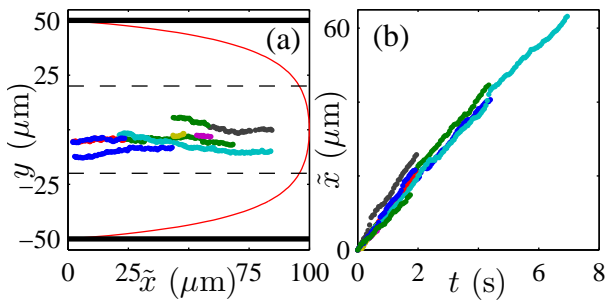}
\caption{{\small (a) Colored symbols: several tracers trajectories in the $x-y$ plane at position $x=5.3$~mm ($\tilde{x}$ is the reduced
position along $x$ within the field of view). The red line shows the theoretical velocity profile at $z=h/2$, the thick lines are the channel edges, and 
the dotted lines indicate the $y$-range of selected trajectories.
(b) Corresponding evolution $\tilde{x}$ vs. $t$ for the trajectories shown in (a). Linear fits (not shown for clarity) yield an estimate of the maximal velocity. }
\label{fig:ProfV}}
\end{center}
\end{figure}

Typical trajectories are shown in Fig.~\ref{fig:ProfV}(a) in the $x-y$ plane, along with the theoretical profile calculated for a rectangular channel with transverse dimensions $h = 35~\mu$m, $w = 100~\mu$m~\cite{Bruus}. 
Note that we neglect here the transverse contribution due to the pervaporation-induced flow across the membrane. Transverse components of the velocity profile are indeed of the order of $q_e/h$ much smaller than the component along $x$ of the order of $v_x(x) \ll x/\tau_e$, see for instance the Supporting Information of Ref.~\cite{Selva:12}. This is actually the same argument which justifies the one dimensional approximation contained in Eqs.~(\ref{eq:v}-\ref{eq:convdiff})~\cite{Schindler:09}.
%%%%%%%%%%%%%%%%%%%%%%%%% 	
%%%%%%%%%%%%%%%%%%%%%%%

We select trajectories with $-20<y<20~\mu$m to minimize
the dispersion due to the Poiseuille shape of the velocity profile (theoretical expected deviation from the maximal velocity $<5\%$), from which we extract the maximal velocity at a position $x$ (size of the field of view $\simeq 100~\mu$m $\ll L$), see Fig.~\ref{fig:ProfV}.
Finally, we estimate the mean velocity $v(x)$ using the relation between the maximal velocity and the average velocity in a rectangular channel $h\times w$~\cite{Bruus}.  
Errors, estimated from the standard deviations over several trajectories, are about $\pm 7\%$, and mainly arise
from the Brownian motion of the tracers which disturbs the measurements of such small velocities 
($\simeq 1$-10~$\mu$m/s).

\section*{Results}
\begin{figure}[htbp]
\begin{center}
\includegraphics{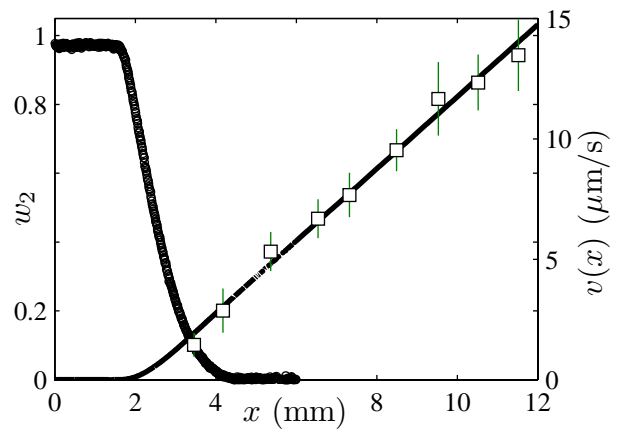}
\caption{{\small Steady concentration profile $w_2(x)$ (black circles).
Squares: velocity $v(x)$ measured using particle tracking velocimetry. The continuous line corresponds to Eq.~(\ref{eq:vint}), see text.  Errorbars for $v(x)$  are estimated from the standard deviations over several trajectories.}
\label{fig:ProfilPhiV}}
\end{center}
\end{figure}

Figure~\ref{fig:ProfilPhiV} displays the main result of our work: combined measurements of the steady concentration gradient $w_2(x)$ and velocities $v(x)$ at several $x$ positions.
%%%%%%%%%%%%%%%%%%%%%%%%% 	
%%%%%%%%%%%%%%%%%%%%%%%

The concentration profile displays a wide plateau at $w_2^\star \simeq 0.97 \pm 0.01$ and 
a decrease to $w_2 \simeq 0$ in the range $x=1.6$--4.3~mm.
The plateau value corresponds to an external humidity $a_e = a(w_2^\star) = 0.09 \pm 0.03$. 
Note that we did not find strictly a null humidity as imposed, 
and this slight mismatch probably comes from mass tranfer resistance within the gas phase,
which results in an imposed humidity $a_e \geq 0$ at the membrane.

Eq.~(\ref{eq:v}) predicts a linear velocity profile far from the
concentrated glycerol region, i.e. in the pure water region. A linear 
fit of $v(x)$ for $x>4.5~$mm leads to $\tau_e = 605~\pm 10~$s.
 The whole velocity profile is then computed using Eq.~(\ref{eq:vint}) from
the measurements $w_2$ vs. $x$, see the continuous line
in Fig.~\ref{fig:ProfilPhiV}.

\begin{figure}[htbp]
\begin{center}
\includegraphics{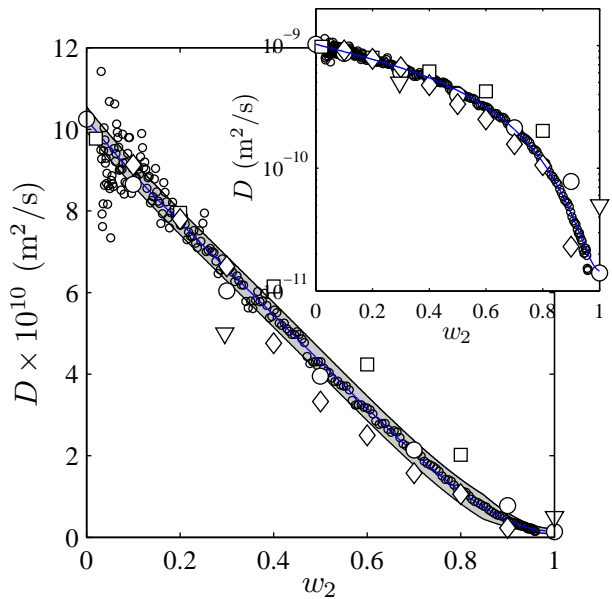}
\caption{{\small Small circles: mutual diffusion coefficient $D$ extracted from the measurements 
displayed in Fig.~\ref{fig:ProfilPhiV} using 
Eq.~(\ref{eq:equil}).
$\circ$ data from Ref.~\cite{DERRICO2004},  
$\square$  from Ref.~\cite{Ternstrom:96},
$\Diamond$  from Ref.~\cite{Nishijima:60},
$\triangledown$ from Ref.~\cite{RASHIDNIA2004}. 
The continuous line is the fit by Eq.~(\ref{Dfit}) and
the gray area displays rough estimates of the error on such measurements, see text.
The inset displays the same data in a semilog plot. }\label{fig:Dphi}}
\end{center}
\end{figure}

Mutual diffusion coefficient is finally estimated from such measurements using 
Eq.~(\ref{eq:equil}). 
Note that the high accuracy of our concentration measurements makes it possible to estimate precisely the numerical derivative $\partial_x w_2$ using a moving average filter 
of width $\delta x = 200~\mu$m only (Matlab functions {\it smooth} then {\it gradient}).  The resulting data are shown in Fig.~\ref{fig:Dphi}.

%%%%%%%%%%%%%%%%%%%%%%%%% 	
%%%%%%%%%%%%%%%%%%%%%%%
$D(w_2)$ shows a significant decrease from $D \simeq 9.8\times10^{-10}$~m$^2$/s at 
$w_2 \simeq 0.02$ to $D \simeq 0.15\times10^{-10}$~m$^2$/s at $w_2 \simeq 0.96$. 
Note also that these measurements are more scattered at low $w_2$, probably due to the difficulty to 
estimate precisely $\partial_x w_2$ in this concentration range. 
Figure~\ref{fig:Dphi} also displays pointwise measurements reported in the literature. The agreement is correct, even with the latest data set~(mean difference $< 0.5\times 10^{-10}$~m$^2$/s) obtained using the Gouy interferometric technique~\cite{DERRICO2004}.
Our data are also well-fitted by:
\begin{eqnarray}
D / (10^{10}~~\text{m}^2/\text{s})&=&10.25-13.08w_2+8.62w_2^2 \notag\\
&&-17.65w_2^3+ 11.98 w_2^4\,,
\label{Dfit}
\end{eqnarray} 
over the range $w_2 = 0.02$--0.96, see the continuous line in Fig.~\ref{fig:Dphi}.  Note that $D(w_2\to 0) = 10.25\times 10^{-10}~$m$^2$/s corresponds to the diffusivity at infinite dilution measured using  
the Taylor dispersion technique~\cite{DERRICO2004}.
The comparison with other measurements from the literature confirms the validity of our approach, but also points out its strength for providing accurate and continuous measurements over a wide range of concentration using a single experiment. 

 A precise estimate of the error for the reported values $D$ vs. $w_2$ is a non-trivial task as our measurements depend on
the calibration accuracy, on the precision of the particle tracking measurements, on the calculation of the numerical derivative $\partial_x w_2$, on the precision of the input parameters $a(w_2)$ and $\rho(w_2)$, but also
on the  precision of the micro-fabrication process (channel geometry). To yield a rough estimate of the error, we assume in the following that the precision over the density and the activity is infinite, and that the transverse dimensions of the channel 
are strictly uniform over the channel length thus leading to a uniform $q_e$. The accuracy of the measurements of $\tau_e$ is high, $\tau_e = 605 \pm 10$~s, and variations of $\tau_e$ of the order of $\pm 10$~s lead to deviations of $D$ of the order of $\pm 1$\% only. Actually, the main errors of the data shown in Fig.~\ref{fig:Dphi} stem 
from the computation of the numerical derivatives $\partial_x w_2$, combined with the absolute precision of the calibration curve ($\pm 0.01$). Using shifted calibration curves by $w_2 \pm 0.01$ and different spans for calculating $\partial_x w_2$, we managed to give an upper and a lower bound to the measurements of $D$, see the gray area plotted in Fig.~\ref{fig:Dphi}. These rough estimates correspond to a typical error of the order of $\pm 5\%$ over the whole concentration range.

\section*{Conclusions and discussions}

The accuracy of our measurements first arises from the outstanding control of the mass transport phenomena at the microfluidic scale~\cite{Squires:05}. Indeed, the small dimensions ensures that the concentration gradient is only governed by a balance between pervaporation-induced convection and molecular diffusion. 
For instance, the unavoidable buoyancy-driven flows induced by the concentration gradient displayed in Fig.~\ref{fig:ProfilPhiV}, associated to a density gradient orthogonal to the gravity, are very small as the latter scales as $v_b \sim h^3$. More accurate estimations using the 
lubrication approximation and using values of the viscosity of glycerol/water mixtures~\cite{Glycerol}, lead to maximal values $v_b$ of the order of 150~nm/s, see for instance Ref.~\cite{Selva:12}. The associated P\'eclet numbers are also extremely small, $\text{Pe} = v_b h /D \leq 0.01$, ensuring that mass transport is indeed governed by molecular diffusion only in such a confined geometry. 
Furthermore, miniaturization combined with the accurate knowledge of the geometry ensured by the microfabrication process ($e$, $h$, $w$) makes it possible to perform a quantitative analysis of mass transport using simple mass balance equations.
The last reason of the high accuracy of our experiments arises from the precision of the 
measurements of $\tau_e$ and $w_2(x)$. Accuracy of the measurement of $\tau_e$ is again 
due to the strict control of hydrodynamic flows at small scales, whereas Raman micro-spectroscopy is a suitable technique for obtaining spatially-resolved concentration profiles with  high precision. 
Any analytical technique which is able to yield absolute concentrations with an accuracy 
of $\simeq 0.01$ at a spatial resolution down to 1--10~$\mu$m would a priori also lead 
to similar results as those shown above. This opens the possibility of using many other analytical
techniques such as fluorescence microscopy, small-angle X-ray scattering SAXS, Fourier Transform InfraRed spectroscopy FTIR, or interferometry, to estimate precisely mutual diffusion coefficients for other liquid binary mixtures.
Note also that the above methodology could be also extended to non-aqueous binary mixtures
using the pervaporation properties of PDMS to other solvents~\cite{zhang16,Ziemecka:15} or
provided that solvent-compatible membranes can be embedded in microfluidic devices, as demonstrated 
for instance by Demko {\it et al.}~\cite{DEMKO2012}.  Finally, one could also 
probably measure mutual diffusion coefficients down to $\sim 10^{-12}$~m$^2$/s 
using the control of the evaporation rate $q_e$  imparted by the tunable geometry ($e$, $w$, $h$),
ensuring that the one dimensional approximation implied in Eqs.~(\ref{eq:v})--(\ref{eq:convdiff}) is still valid~\cite{Schindler:09}.

Note that our technique, based on spatially-resolved measurements of concentration profiles in a pervaporation process, is reminiscent of other techniques which exploit similar mechanisms, as for instance solvent evaporation from polymeric coatings~\cite{Siebel:15}. Such experiments also lead to precise estimates of $D$  as a function of the solvent concentration using a single experiment (with possibly strong variations of $D$), but again from time-resolved measurements, whereas our technique provides a {\it steady} out-of-equilibrium regime.
We thus hope that the methodology detailed above will be used to measure accurate values of mutual diffusion coefficients
in many other binary liquid mixtures, including complex fluids such as polymer solutions, for which classical techniques are either inadequate or tedious. 
%%%%%%%

\begin{acknowledgements}
The authors thank Jacques Leng 
for useful discussions, and Agence Nationale de la Recherche, ANR EVAPEC (13-BS09-0010-01) for funding.
\end{acknowledgements}

%\bibliography{C:/Users/salmon/Documents/ARTICLE/BIB/onion}

\end{document}